\documentclass[preprint,12pt]{elsarticle}

\usepackage{graphicx}
\usepackage[caption=false]{subfig}
\usepackage[colorinlistoftodos]{todonotes}
\usepackage[colorlinks=true, allcolors=blue]{hyperref}
\usepackage{amssymb}
\usepackage{natbib}
\setcitestyle{square, comma, numbers,sort&compress, super}

\journal{Materials Characterization}

\begin{document}

\begin{frontmatter}


\title{Advanced data mining in field ion microscopy}

\author[First]{Shyam Katnagallu}
\ead{s.katnagallu@mpie.de}

\author[First]{Baptiste Gault\corref{cor1}}
\ead{b.gault@mpie.de}

\author[Second]{Blazej Grabowski}
\ead{b.grabowski@mpie.de}

\author[Second]{J\"org Neugebauer}
\ead{j.neugebauer@mpie.de}

\author[First]{Dierk Raabe}
\ead{raabe@mpie.de}

\author[Second]{Ali Nematollahi\corref{cor1}}
\ead{a.nematollahi@mpie.de}

\cortext[cor1]{Corresponding Author}

\address[First]{Department of Microstructure Physics and Alloy Design, Max Planck Insitut f\"ur Eisenforschung }
\address[Second]{Department of Computational Material Science, Max Planck Insitut f\"ur Eisenforschung.}

\begin{abstract}
Field ion microscopy (FIM) allows to image individual surface atoms by exploiting the effect of an intense electric field. Widespread use of atomic resolution imaging by FIM has been hampered by a lack of efficient image processing/data extraction tools. Recent advances in imaging and data mining techniques have renewed the interest in using FIM in conjunction with automated detection of atoms and lattice defects for materials characterization. After a brief overview of existing routines, we review the use of machine learning (ML) approaches for data extraction with the aim to catalyze new data-driven insights into high electrical field physics. Apart from exploring various supervised and unsupervised ML algorithms in this context, we also employ advanced image processing routines for data extraction from large sets of FIM images. The outcomes and limitations of such routines are discussed, and we conclude with the possible application of energy minimization schemes to the extracted point clouds as a way of improving the spatial resolution of FIM.
\end{abstract}


\end{frontmatter}

\section{Introduction}
\label{S:1}

Field ion microscopy (FIM), invented in 1951 by Erwin M{\"u}ller \cite{muller1955ergebn,muller1951Feldionenmikroskop,Muller1956}, is a high electric field technique which uniquely enables imaging of surfaces with atomic resolution. FIM is based on ionization of an imaging gas in the vicinity of a field-emitter tip as a consequence of the locally high electric field. The high electric field is achieved by applying a high voltage of a few kilovolts onto a very sharp needle-shaped specimen maintained at a temperature usually below 80 K. Specimens are either electropolished \cite{miller2014atom} or milled with a focused ion beam (FIB) \cite{prosa2017modern} into a very sharp needle tip with an end radius below 100 nm. The advantage of using FIB for specimen preparation lies in its site specific application for extracting tips in microstructure regions of high interest such as across internal interfaces. An excellent review on using FIB for site specific specimen preparation can be found in reference \cite{prosa2017modern}.

Once the specimen is mounted an imaging gas is introduced. The introduced imaging gas gets  attracted by the cold surface due to polarization forces. The gas atoms then thermally accommodate with the cold tip surface by performing a series of ''hops''. Surrounding the tip surface there exists a critical surface, where the maximum ionization occurs. This surface usually lies around 1-4 \AA \  above the tip \cite{de1986best}. During the ''hops'', the ionization probability for the gas atoms can be considerable as they spend a significant amount of time in the critical surface. As a consequence an electron can tunnel from the imaging gas atom into the tip. The ionized gas atom is accelerated away from the positively biased tip and towards the detector, where gas ions contribute to image formation.

The surface is the intersection of the crystalline lattice with the imposed end shape, often approximated as nearly-spherical. Owing to the discreteness of the atomic arrangement at these scales the tip apex curvature is, in reality, made of atomic scale crystallographic terrace features where some of the top, edge and corner atoms are naturally protruding, producing local field enhancement. This means that these exposed atomic terrace positions are sites of magnified electrostatic field strength and also of aberrated field direction. The amount of gas atoms ionizing depends on such a local enhancement of the electric field. This variation in electric field strength across the surface atoms gives the final contrast in the FIM image. Overall the contrast in FIM also depends on the gas supply function and adsorption behavior \cite{Chen1971a,Schmidt1993,tsong2005atom}.  Atomic resolution can be attained in some cases on certain high index facets where the surface field distribution is corrugated enough to give contrast in the image. By collecting the gas ions on a phosphor screen, an image is formed that reveals the distribution of the electrostatic field near the surface and the current created by the number of incoming imaging gas ions.

Historically, images were collected on a film in a dark room after sufficient exposure on the screen was achieved by field ionization \cite{muller1951Feldionenmikroskop}. Eventually,  field ion microscopes were fitted with a stack of micro-channel plates (MCP) in-front of the phosphor screen. The image on the phosphor screen can be recorded with a high-resolution high-frame-rate camera. The MCPs act as a photo multiplier by creating an electron cascade in response to the gas ion impact. Another variant of FIM is referred to as eFIM$^{\rm TM}$, which is performed on a local electrode atom probe (LEAP) \cite{larson2013local} using the delay-line detector used in the atom probe mode.

The atomic resolution of FIM made it a popular technique for studying internal interfaces \cite{Brandon1964,Ranganathan1966,Deconihout1994} and dislocations \cite{Smith2013,Smith1968,Taunt} at unprecedented atomic positioning resolution. When exposing the tip not only to the minimum field strength required for ionizing the imaging gas but also for evaporating the tip atoms themselves continually, the method is rendered depth sensitive. This means that the specimen can be investigated tomographically along the tip longitudinal axis, which has led to the development of 3DFIM \cite{Vurpillot2007}. The emergence of atom probe tomography (APT), which is additionally able to characterize the chemical identity of the imaged atoms, lead to some decline in the usage of the FIM technique by the materials science community. Nevertheless 3DFIM offers the important advantage over APT of a significantly higher spatial resolution with 100 \% positional detection efficiency in 3 dimensions in some cases. This high degree in positional accuracy allows characterization even down to single point defects in a material, a feature not offered by any other device. This causes currently an increased interest in the 3DFIM technique.

Ultimately, owing to all the advances in detector technology, 3DFIM is capable of producing large and accurate tomographic datasets containing information on sequential atomic positions. These large datasets lead to a new tremendous challenge of how to manage the data. Presently, there is a lack of efficient data handling and data treatment algorithms to extract pertinent information from these datasets in an (a) automated; (b) fast; (c) user-independent; (d) and error quantified manner. For instance, characterization of a volume of 0.001 $ \mu \textrm{m}^3 $ (a typical sample size analyzed in 3DFIM) produces in the range of $ 2\times10^{15} $ images (assuming a constant field evaporation rate and capture speed). Hence, there is a great need for efficient algorithms and data mining routines to fully exploit the potential of 3DFIM. To this end, M. Dagan et al. \cite{Dagan2017} have proposed an atom by atom data extraction routine for reconstructing 3DFIM data. Building on their work we recently proposed a new method to extract atomic positions from 3DFIM datasets \cite{Katnagallu2017}. With this article, we focus our attention on using various modern image processing and machine learning algorithms for extracting information from 3DFIM.

The developments in Artificial Intelligence (AI), especially in computer vision  have been explosive. Modern machine learning algorithms enable a fully automated detection and classification of objects in a picture. We give an overview about these advanced data mining tools and how they can be utilized to extract the wealth of information from 3DFIM images. We have implemented some of these concepts within a set of routines in Python\textsuperscript{TM} (Python Software Foundation; Python Language Reference, version 2.7; Available at http://www.python.org) employing the SciPy package \cite{jones2014scipy}. These routines allow us to extract the relevant information from a large number (order of several 10,000s) of FIM images in very short computational times (order of minutes).  

\section{Existing data extraction routines for 3D field ion microscopy}

A 3DFIM experiment produces a series of images of the continually field evaporating surface. A main challenge lies not in acquiring such large datasets but rather in analyzing them. The article by Vurpillot et al. \cite{vurpillot2017true} serves as an excellent review for the current state-of-the-art and the main issues associated with data extraction from large FIM datasets. We briefly review here the available analysis methods and also some recent developments. Broadly speaking the analysis methods can be categorized into an atom by atom approach and a geometrical approach. \par 

The first approach towards advanced 3DFIM analysis was already developed in the early 70s for characterizing radiation damage in metals at the atomic scale  \cite{Scanlan1971,Beavan}. In these early approaches FIM images were captured on film which later were developed and manually analyzed individually. The captured FIM images were dissected atom-by-atom and the positions of atoms and defects were marked manually. Owing to the associated cumbersome analysis methods, systematic FIM studies of more complex atomic scenarios remained an exception. Taking the additional disadvantage of 3DFIM of being insensitive to the chemical nature, APT became gradually the more dominant technique. Yet, FIM's ability to characterize atomistic defects such as vacancies in three dimensions is still unparalleled with any other technique. In this context the drastic increase in computing power became an essential asset when Dagan et al. developed an automated method to reconstruct 3DFIM data atom by atom \cite{vurpillot2017true,Dagan2015}. The algorithm takes advantage of layer by layer evaporation and the atoms are identified based on a threshold intensity. The final coordinates are converted to real space based on theoretical nearest neighbor distances. This work lead to a rise in interest around the physics of image formation in FIM and also in the use of the associated computationally enhanced analysis techniques. \par
The geometrical approach to 3DFIM atomic position reconstruction introduced by Vurpillot et al. consists in stacking the digital images obtained from a 3DFIM experiment \cite{Vurpillot2007}. The image stack is then corrected, assuming a known projection law, a specimen's geometry and a constant evaporation rate. The stacking approach does not provide atomistic precision but is rather used for investigation of segregation, clustering and fine scale precipitation studies \cite{Cerezo1992,Cazottes,Jessner,danoix2012atomic,Akre2009}. This method can also be used to identify crystallographic planes and dislocations which are hard to spot in a 2DFIM image.

Both the atom by atom reconstruction approach and the geometrical method suffer from their own limitations. For instance, the atom-by-atom approach is limited to regions with atomic resolution, and thus a 3D reconstruction is only possible around certain high index facets. The geometrical approach looses the atomic positioning precision due to the simplistic geometrical assumptions of the tip shape. In the following sections we showcase how various data extraction methods can be employed to further improve the atom-by-atom analysis approach and recover as much positioning information as possible. In addition, the use of machine learning algorithms to extract the physics behind field ionization and evaporation is also explored.\par

\section{Supervised and unsupervised machine learning}

Machine learning (ML) algorithms are currently exploited to derive systematic insights from very rich experimental datasets and  for solving complex problems in various disciplines \cite{mjolsness2001machine}. Progress in  ML has lead to decision rules that can in some cases be automatically derived by specific algorithms that are capable of learning, whilst exploiting the speed and the robustness of the available advanced computer infrastructure. \par

Machine-learning methods can be grouped into two major categories depending on the approach to a given problem viz. supervised and unsupervised learning \cite{huang2006kernel}. Supervised learning algorithms try to identify the relationship between input and output. This dependency is learned as a function $ f(x) $ by using a set of labeled data \{${X = [a_i,b_i],\,i = 1,...,N}$\} consisting of $N$ pairs $(a_1,b_1)$, $(a_2,b_2),\,...\,,(a_N,b_N)$, where the input variables $ a_i $ are $D$-dimensional vectors $a_i \in R^D$ and the output variables (or system responses) $ b_i $ are discrete values (e.g., Boolean) for classification problems and continuous values ($b \in R$) for regression tasks. Support Vector Machines (SVMs) and Artificial Neural Networks (ANN) are widely used techniques that fall in this category. Typical tasks that can currently routinely be carried out by a supervised machine learning algorithm are image segmentation and classification. In the computer vision community, semantic segmentation, which is  an extremely challenging task, aims to partition the image into semantically meaningful parts (such as differentiating a cat, a car or Einstein in the same image), and to classify each part into one of the predetermined classes making use of ML or the more advanced deep learning methods \cite{garcia2017review}. 3DFIM represents a very good example of a complex problem where rather large experimental datasets are available and which are to date widely under-exploited. For the case of FIM data analysis, this concept could help in determining whether a group of pixels represents, e.g. an atom, a defect such as a vacancy, a dislocation or a grain boundary. This idea can be also extended to 3DFIM, where the data can be subjected to systematic segmentation and classification into sub-volumes that may contain or represent these various features. The motivation for using such algorithms lies in enabling the automated reconstruction of a fully atomically-resolved three dimensional point cloud representing the imaged specimen, including its population of crystalline defects. However, to apply deep learning algorithms for semantic segmentation, the  structure of 3DFIM data must be defined precisely, followed by labeling of a large enough dataset to be used as a training dataset for the algorithm. 

Unsupervised learning refers to any ML process that attempts to learn the structure in an unlabeled dataset \{${X = [a_i],\,i = 1,...,N}$\}, where $a_i \in  R^D$ in the absence of the output variables $ b_i $ \cite{huang2006kernel}. Various clustering techniques and dimensionality reduction routines fall into this category. The fact that the unsupervised learning techniques do not need a labeled training dataset is a considerable advantage. However, since there is usually no sufficiently robust or large dataset in this field to train the algorithm reliably enough, accuracy that can be obtained by these methods is usually lower than that achieved by the help of supervised learning algorithms. In this paper, we focus on unsupervised methods to identify the underlying structure of the FIM data, to visualize them and to label them for further application of supervised learning techniques. \par
\begin{figure}
\centering
\includegraphics[width=0.9\linewidth]{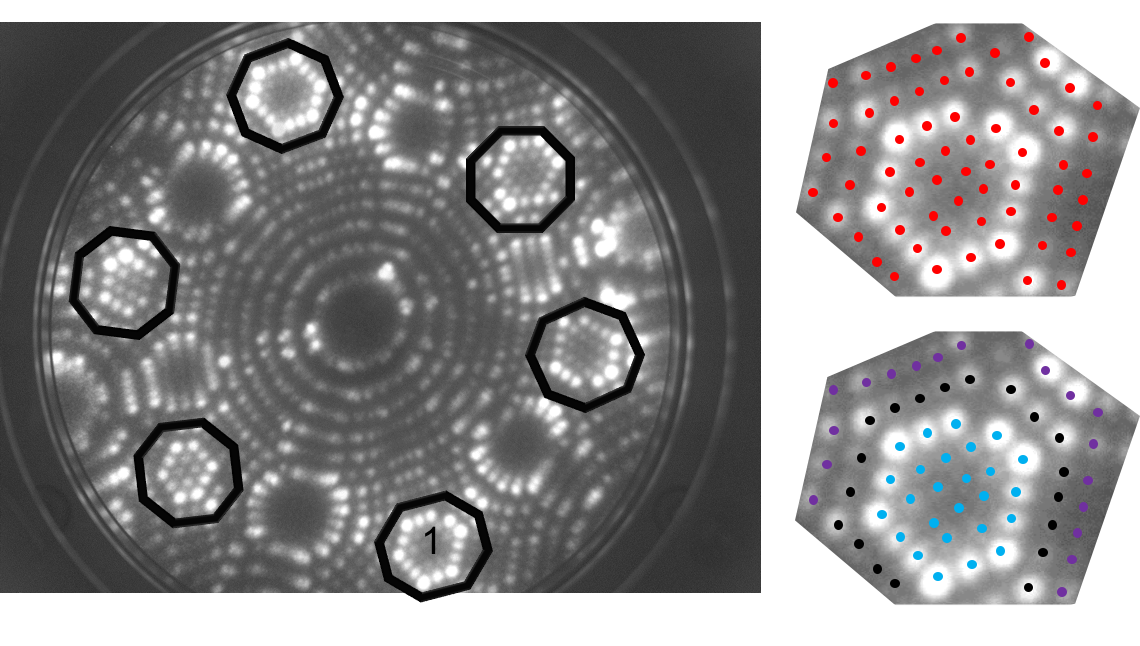}
\caption{FIM image of pure bcc W obtained using He as imaging gas. Regions with atomic resolution are highlighted. The region marked by 1, is used for identifying all the atoms (marked in red in the upper right corner). Machine learning is then used to assign correct labels to those atoms that assemble crystallographic planes. In this case, atoms labeled blue belong to the first, black to the second and purple to the third plane layer, respectively, assembling altogether a pyramid type motif (lower right corner).}
\label{fig:outline}
\end{figure}

\section{Advanced data mining routines for 3D field ion microscopy}
In this section we describe the advanced data mining routines which include unsupervised learning routines (Sec.~\ref{ML}) and a routine purely based on image processing (Sec.~\ref{image}). To demonstrate the utility of these methods we apply them to a 3DFIM dataset obtained for bcc tungsten. The specimen for 3DFIM was obtained from a tungsten wire oriented along the [011] direction. A very sharp needle like shape was achieved by using electrochemical polishing at 5-8 V using AC in a 5\% molar NaOH solution. The local radius across the main [011] pole was estimated to be approximately 20 nm by the ring counting method \cite{drechsler1960analyse}. 3DFIM experiments were performed on a 3DAP-LAR FIM \cite{Panayi:2006:3DAP:LAR:FIM}. FIM images were obtained on a phosphor screen, and recorded using a CCD camera (AVT Stingray) at 15 fps rate. The resolution of each recorded image was set to 1280 pixels $\times$ 960 pixels. From this complete image, a region of interest (ROI) was selected around the (222) crystallographic plane. Each image of the ROI is $150 \times 150$ pixels in size. A set of (222) planes was chosen for the data analysis as the atomic density of these planes in a body centered cubic structure is sufficiently low to produce atomic resolution in FIM. In the following, ''FIM image'' generally refers to the selected ROI image of $150 \times 150$ pixels.

\subsection{Unsupervised machine learning applied to 3DFIM\label{ML}}
Here we use unsupervised machine learning to understand the underlying structure of 3DFIM data and use this information to test supervised learning algorithms for data extraction. It will be seen that the extracted structure of 3DFIM data is related to various field evaporation phenomena. Although these phenomena of field evaporation have been long established, the ability of a machine to decipher this structure from 3DFIM data serves as a proof of concept.
\subsubsection{Dimensionality reduction}
\label{sec:Dimensionality reduction}

In problems with high levels of complexity, one of the challenges lies in extracting information from a dataset that is made of a large number of samples and where each sample has high-dimensionality. The 3DFIM dataset used herein contains 10,000 FIM images each with an image size of $ 150\times150 $ pixels, i.e. 22,500 dimensions (considering only the ROI). Relevant extraction of information from such data requires identifying pertinent variables, determining the interaction of these variables with each other, and then reformulating the data using exclusively these specific variables. This procedure is referred to as "dimensionality reduction" and simplifies any further processing. Mathematical hurdles of working with high-dimensional data are often called the “curse of dimensionality” \cite{friedman2001elements}. When the dimensionality of the data increases, a good representation of the data in 2D or 3D Euclidean spaces can be very helpful to reveal various phenomena present in the data but such a representation might not always be possible \cite{lee2007nonlinear}. \par

For data with high dimensionality, two complementary strategies are employed to either avoid or at least reduce issues related to dimensionality. First, there are often a number of input variables that do not or only weakly affect the output and can hence be considered as irrelevant and ignored. Second, the dependencies amongst the pertinent variables are established. Once the relevant variables have been established, the dimensionality of the observed data tends to be still larger than necessary. Let us consider two highly correlated variables where information about one can be derived from the other. Correlations between such variables can be sometimes very complex and retaining only one of them might not be sufficient to encompass the whole information. Hence, instead of arbitrarily removing a variable from the pair, the best is to reduce the dimensionality of the data by trying to find a new set of transformed variables that keeps as much of the available information as possible. \par

The new set of variables must possess the following features: It must 1)~contain fewer variables than the original set and 2) preserve the most important information contained in the initial dataset. We hence seek a transformation that contains and shows the same information from a different perspective, i.e. a projection which preserves the geometry whilst representing the relevant objects \cite{weinberger2006unsupervised}. Linear and nonlinear transformations of the input used herein are referred to as projections, mainly because all these transformations aspire to preserve the characteristics that are geometrical or have a geometrical interpretation. In other words, when there is a dependency amongst two or more variables, their joint distribution does not spread over the whole space. The dependencies in the data create a structure in the distribution, that can be seen in the form of a geometrical locus. Dimensionality reduction attempts to eliminate any redundancies in the initial set of variables. Linear multivariate analyses such as principal component analysis (PCA) or multidimensional scaling (MDS) have long been used to find such projections, but cannot always reveal low-dimensionality structures when the manifold formed by the geometrical loci of data is curved \cite{Vlachos2017}. Hence, there has been recent interest in algorithms identifying non-linear manifolds in data \cite{tenenbaum2000global,lee2007nonlinear,weinberger2006unsupervised}.
There are many, somewhat heuristic, methods to discover non-linear manifolds based on the preservation of the geometric properties of local neighborhoods within the data, while embedding, unfolding or otherwise projecting the data to occupy fewer dimensions. Algorithms such as Isometric feature mapping (Isomap) or maximum variance unfolding (MVU) attempt to preserve local distances. They estimate global manifold properties by continuation across neighborhoods and then project the manifold to lower dimensions by classical methods such as PCA or MDS. In the following, we discuss the application of a linear method (PCA) and a non-linear method (Isomap) to the 3DFIM dataset.

\subsubsection*{Principal Component Analysis}
PCA is a popular technique for dimensionality reduction that aims to identify the most consequential basis to re-express a given dataset. The new basis should reveal the underlying structure in the dataset and filter out the noise \cite{lee2007nonlinear}. Mathematically, we can define PCA as follows: For  $ N $ high dimensional input variables \{${X = [a_i],\,i = 1,...,N}$\}, where $a_i \in  R^D$, PCA is employed to find a linear subspace with lower dimensionality $ d $ ($ d\leq D $), such that the maximum variance is maintained in the subspace. The linear subspace can be defined by $ d $ orthogonal vectors also called principal components, say $ u_1,u_2,...,u_d $, forming a new coordinate system. Ideally $d \ll D $ (worst case would be $ d=D $). In other words, PCA aims to reduce the dimensionality of the data, while preserving its information content which comes from the variation (or, equivalently, minimizing the loss of information). The principal components are given by the top eigenvectors of the $ D\times D $ covariance matrix $ C =\frac{1}{N}\sum_{i}X_i\,\cdot X_i^T $, with $i$ running over the rows of the matrix $X$ \cite{lee2007nonlinear}.

\begin{figure}[!t]
\centering
\includegraphics[width=0.75\linewidth]{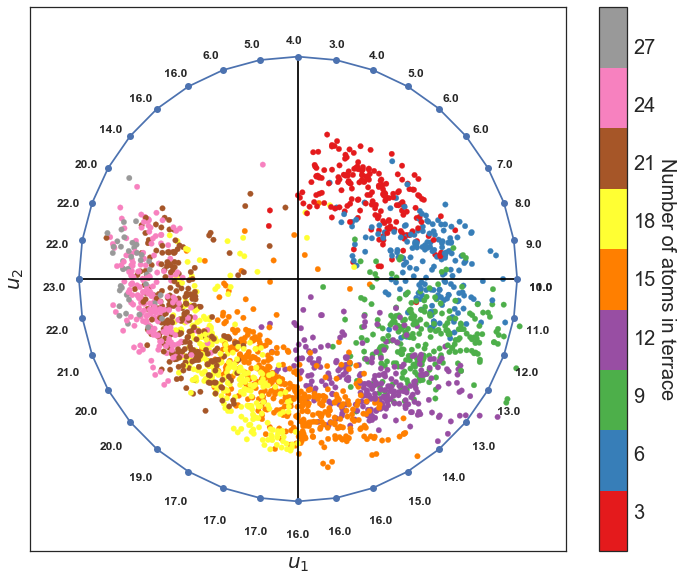}
\caption{PCA applied to the 3DFIM dataset. Each point represents one FIM picture in this dataset, and its color indicates the number of atoms in the first terrace of the corresponding picture. The numbers on the outer circle represent the averages over the number of atoms in the first terrace for that particular arc of the circle.}
\label{fig:PCA_1}
\end{figure}

Figure \ref{fig:PCA_1} shows the result of projecting our 3DFIM dataset onto its first two principal components. In this plot, each point represents one micrograph (with a dimensionality of $ 150\times 150 $), and is color-coded according to the number of atoms in the first terrace of the corresponding picture (see Fig.~\ref{fig:outline}). The numbers on the outer circle represent the average number of atoms on the first terrace of images in the corresponding slice of the circle. The PCA result reveals a cyclic pattern in the 3DFIM dataset and a specific grouping of the colors. Note that starting with the gray color cloud to the left representing a terrace of 27 atoms and going counter-clockwise the number of atoms in the terrace gradually decreases until 3, before a jump to the next terrace occurs.  

\subsubsection*{Isomap}
Isomap is a nonlinear dimensionality reduction method that is built on a linear dimensionality reduction approach, more precisely it is built upon a classical multidimensional scaling (MDS) (see reference \cite{cox2008multidimensional} for details). In MDS, the input $ X $ is projected into a smaller subspace with dimension $d$ by preserving the pairwise Euclidean distances $ |X_i-X_j|^2$, or the dot products $ X_i\,\cdot X_j $, i.e. the $ L^2 $ norm. The orthogonal vectors, $ w_1,w_2,...,w_d $, which form the subspace are the top $ d $ eigenvectors of the $ N \times N $ Gram matrix with elements $ G_{ij} = X_i\,\cdot X_j $ \cite{weinberger2006unsupervised}. Isomap also tries to preserve the local geometry of the data. To this end, one assumes that the high-dimensional data in $ X $ lie on a low-dimensional manifold $ M $, and one replaces the Euclidean distances between the points in the high-dimensional space with the distances on this manifold. Since the manifold is unknown, to a first approximation, a $k$-nearest neighbor graph between points in the dataset is considered, where $k$ needs to be chosen such that any two data points in the dataset are connected by at least one path but it should not become too large to have a computationally efficient approach. The shortest path between the points is then used to approximate the distance on the manifold.

The complete Isomap algorithm has three main steps \cite{weinberger2006unsupervised}:
\begin{enumerate}
\item Constructing a $k$-nearest neighbor graph based on the distances $ d_X (i, j) $ between pairs of points $ i,j $ in the input space $ X $. Either all the points within some fixed radius are connected to each point, or each point is connected with each of its $k$-nearest neighbors. These neighborhood relations are expressed as a weighted graph $ G $ over all the data points, where edges are represented as weights $ d_X (i, j) $ between the neighboring points. 
\item The geodesic distances $ d_{M (i, j)} $ between all pairs of points on the manifold M are estimated by Isomap from computing their shortest path distances $ d_{G (i, j)}$ in the graph $ G $.
\item Finally a MDS is applied to a matrix of graph distances $ D_G = \{d_G (i, j) \}$, which embeds the data in a $d$-dimensional Euclidean space $  Y $ which preserves the manifold's estimated intrinsic geometry. The coordinate vectors $y_i$ for points in $ Y $ are chosen to minimize the cost function:

\begin{equation}
E = |\tau(D_G) - \tau(D_Y) |_{L^2}
\label{eq:Isomap_cost}
\end{equation}

where $ D_Y $ denotes the matrix of Euclidean distances $ {d_Y(i,j)  = |y_i - y_j|} $ and  $ |A|_{L^2} $ the $ L^2 $ matrix norm $ \sqrt{\sum_{i,j}A^2_{ij}} $ . Distances are converted into the inner products by  the $ \tau $ operator, these characterize the geometry of the data in a form that supports efficient optimization. The global minimum of Eq. (\ref{eq:Isomap_cost}) is achieved by setting the coordinates $y_i$ to the top $d$ eigenvectors of the matrix $ D_G $.

\end{enumerate}

\begin{figure}[t!]
\centering
\includegraphics[width=0.8\linewidth]{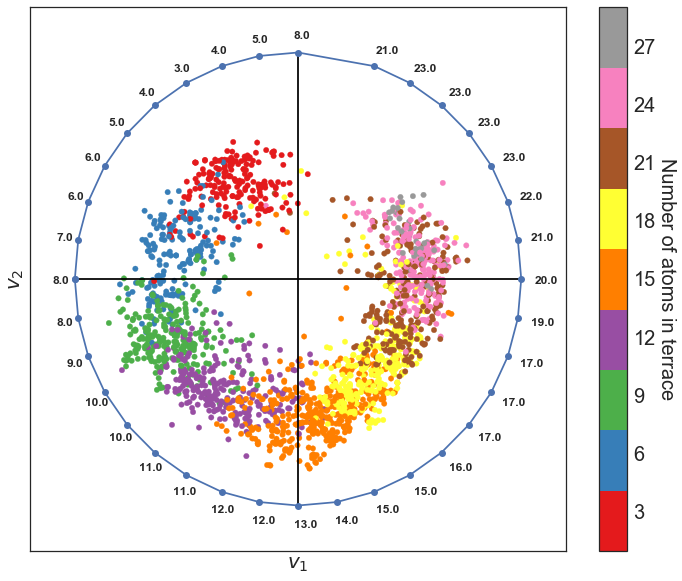}
\caption{Isomap manifold learning on the 3DFIM dataset. Each point represents one picture in the reduced dimensional space, and its color indicates the number of atoms in the first terrace of the corresponding picture. The numbers on the outer circle represent the averages over the number of atoms in the first terrace of pictures.}
\label{fig:Iso_map_1}
\end{figure}

\begin{figure}[t!]
\centering
\includegraphics[width=1\linewidth]{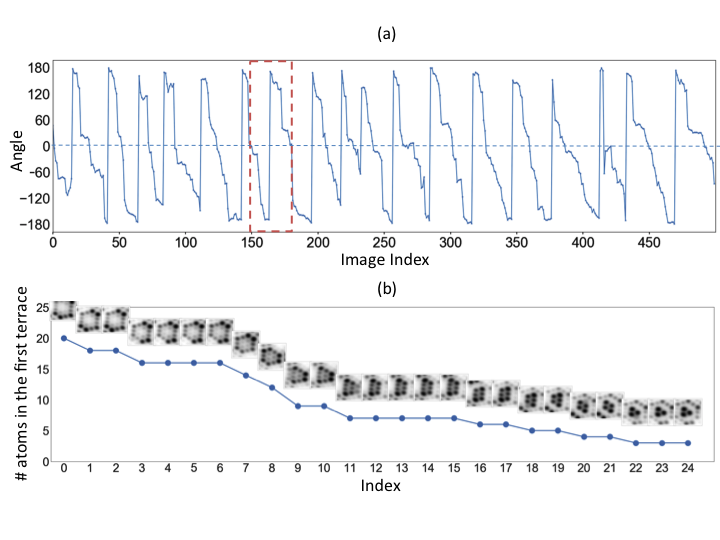}
\caption{(a) Angles of the first 500 3DFIM images after projection onto the Isomap space with horizontal axes (see Fig. \ref{fig:Iso_map_1}). (b) Evolution of a terrace seen from FIM pictures and the number of atoms in the first terrace of the highlighted period  in (a) (red dashed rectangle).}
\label{fig:Iso_map_4}
\end{figure}

Figure \ref{fig:Iso_map_1} shows the result of applying Isomap to the 3DFIM dataset. Each point represents one micrograph (with dimensionality of $ 150\times 150 $) in a reduced 2-dimensional space. Each image is color-coded according to the number of atoms in the first terrace. Again, the numbers on the outer circle represent the averages over the number of atoms in the first terrace of pictures whose Isomap representations are in a slice of the circle with a corresponding angle. The Isomap result reveals a similar cyclic behavior in the dataset as found with PCA. This cyclic behavior detected by both linear and non linear dimensionality reduction algorithms is a consequence of the field evaporation behavior in crystalline materials. Field evaporation in many pure and crystalline materials occurs layer-by-layer and proceeds from atoms sitting at the ledge of a terrace and towards the center. When a terrace field evaporates the atoms sitting on the ledge disappear decreasing the number of atoms on the terrace as the evaporation proceeds. As shown in Fig. \ref{fig:Iso_map_4}(b), by evaporating atoms from the surface of the sample during FIM, the first terrace area decreases until it evaporates completely and the subsequent terrace is exposed. This field evaporation behavior is seen for all terraces. The number of cycles is a measure of the number of atomic terraces evaporated/analyzed during the 3DFIM.

Figure \ref{fig:Iso_map_4} (a) shows the angles of the first 500 3DFIM images after projection onto the Isomap space with horizontal axes (see Fig. \ref{fig:Iso_map_1}). Each period of this plot represents the field evaporation of one terrace from the moment of its exposure to the surface until its complete evaporation. Figure \ref{fig:Iso_map_4} (b) depicts the evolution of the FIM pictures and the number of atoms in the first terrace during one of these periods (highlighted in Fig. \ref{fig:Iso_map_4}(a) by the red dashed rectangle).

\begin{figure}[t!]
\centering
\includegraphics[width=0.8\linewidth]{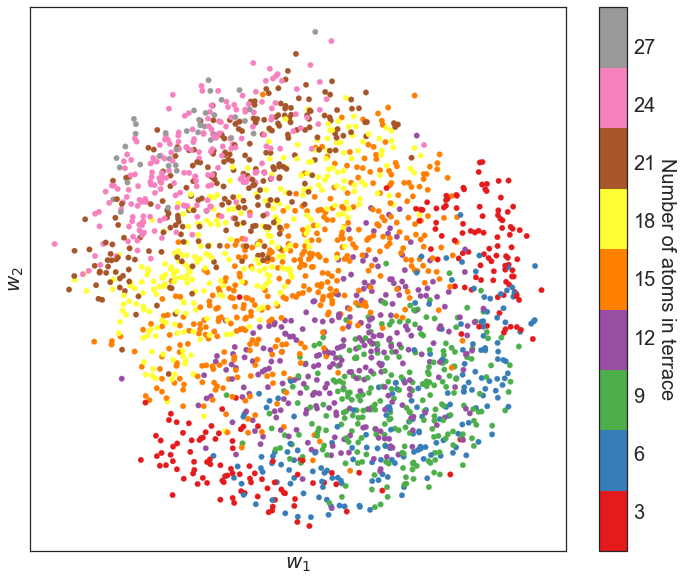}
\caption{MDS applied to the 3DFIM dataset. Each point represents one picture in the reduced dimensional space, and its color indicates the number of atoms in the first terrace of the corresponding FIM picture.}
\label{fig:MDS_1}
\end{figure}

Although Isomap is built on an MDS approach, the MDS algorithm applied to the same dataset is not able to reveal the cyclic pattern observed within the data (see Fig. \ref{fig:MDS_1}). The reason is that the objective function in MDS aims to preserve the pairwise Euclidean distances in the low-dimensional space and therefore it fails to appropriately represent the 3DFIM dataset. A comparison between the projections obtained from Isomap and PCA shows that the Isomap projection reveals finer details and enables a better separation between micrographs based on the number of atoms in the first terraces (see the color of points in Fig. \ref{fig:Iso_map_1}). However, there is still an overlap of colors especially in the images with a higher number of atoms in the first terrace. The reason is that for high numbers of atoms in the first terrace, the number of available configurations that these atoms can adopt is significant. Hence an ideal projection would be one that could distinguish the data based on the configuration of atoms in the corresponding image. To this end, the local information within the images should also be taken into account whilst calculating the similarity (e.g. pair distances) between images. This could be achieved by using autoencoders in combination with convolutional neural networks, which is beyond the scope of the current article. (See reference \cite{goodfellow2016autoencoders} for additional details.)

We showed here that a complex 3DFIM dataset can be used as input for dimensionality reduction efficiently unraveling the underlying structure of the data. Further we discussed that the structure revealed by the linear and non-linear dimensionality reduction is in fact a consequence of the field evaporation behavior of a pure and crystalline material. The output of these algorithms was able to classify the data according to the number of atoms on the first terrace and also, to some extent, the configurations formed by those atoms. Further analyses can be performed in the direction of analyzing the stability of certain configurations for a given crystallographic plane. In the following, we describe some advanced image processing methods and clustering algorithms which can be used to maximize the data extracted from 3DFIM.  

\subsection{Image processing based data extraction routines applied to 3D field ion microscopy\label{image}}

\begin{figure}
\centering
\includegraphics[width=0.9\textwidth]{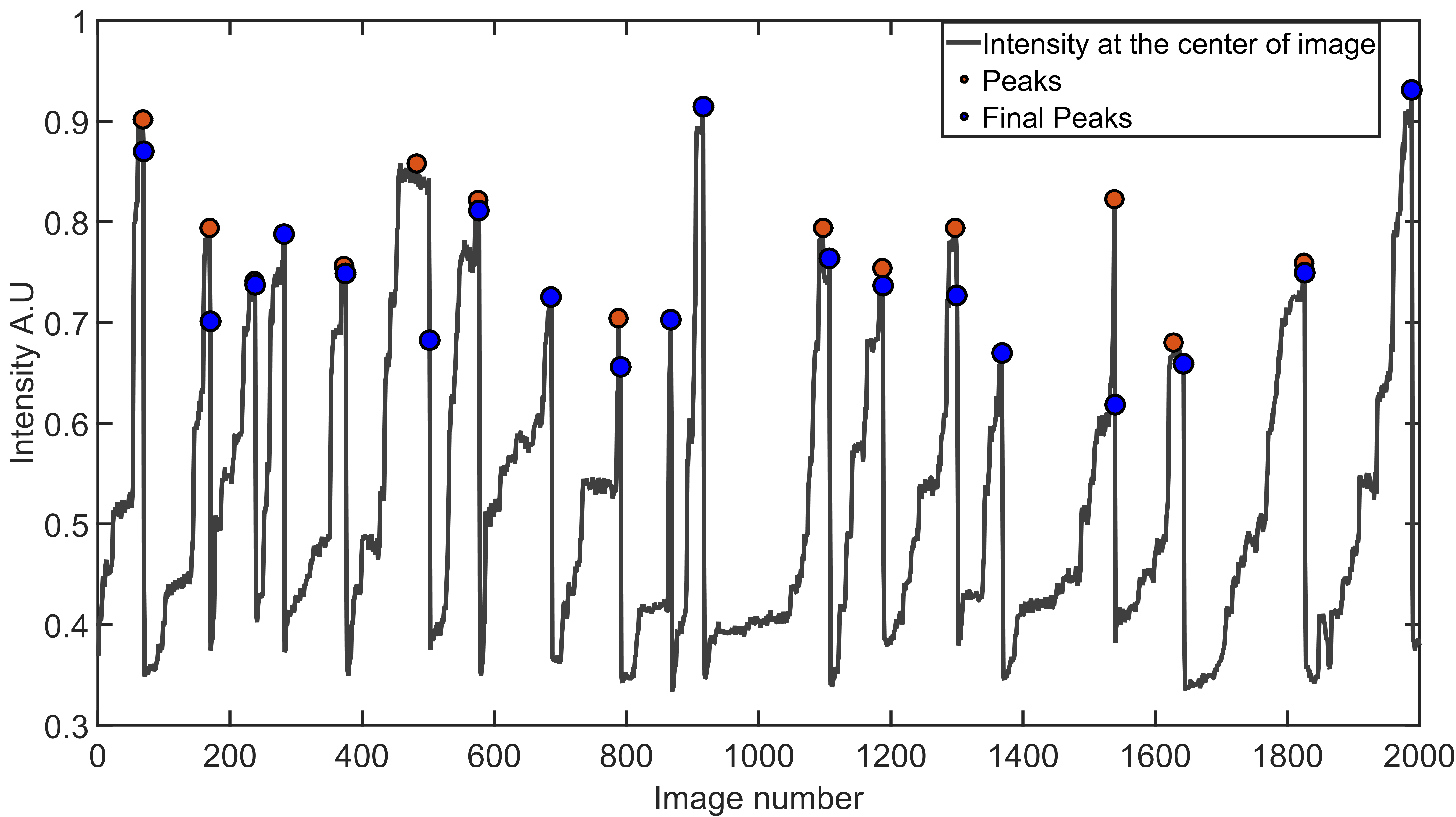}
\caption{Intensity at the center of the FIM image versus the image number. Peaks indicate the number of planes evaporated. Compare with Fig. \ref{fig:Iso_map_4} which also shows the same periodic behavior, a consequence of layer-by-layer evaporation.}
\label{planeEvap}
\end{figure}

In a recent paper \cite{Katnagallu2017}, we showed an improved method for reconstructing 3DFIM data. It was also shown recently [Katnagallu et al., 2017 communicated] that the higher the intensity of an imaged position, the less representative is the imaged position of the true atomic position, which implies that the positions of atoms imaged in the center of a terrace are more accurate. Keeping this in mind, we developed another new routine for 3DFIM data reconstruction. In every 3DFIM data extraction routine, the first step is to determine the overall depth of the analyzed volume. In an atom-by-atom approach, this problem boils down to calculating the number of planes of interest that have evaporated, which can be directly derived from monitoring the intensity in the center of the image. As the evaporation proceeds from the edge of the terrace inwards, the intensity in the center keeps increasing. As the terrace collapses, the atoms left on the terrace are imaged brightly. The reason is the high local curvature leading to an increase in the local field, thus increasing the current of gas ions originating from above of these specific atoms. Such an analysis is shown in Fig. \ref{planeEvap} revealing a clear periodic behavior where each peak indicates the final image of a terrace before complete evaporation. The same information can be obtained from an ML approach as was shown by the dimensionality reduction using PCA and Isomap (see  Fig. \ref{fig:Iso_map_4}(a)).  The number of peaks in this intensity variation thus gives the number of planes evaporated. 

After this number is computed, the image at which a new plane is exposed can also be calculated based on the peak positions. The determination of the depth of the data enables the next steps of the data extraction procedure. These steps can be classified into three categories a) identification of atoms (atom detection; Sec.~\ref{atom}) b) assigning the right crystallographic plane to the identified atoms (plane detection; Sec.~\ref{plane}) and c) depth assignment based on the number of planes evaporated. Depth assignment is easier to implement and as described above capitalizes on layer-by-layer evaporation, while the first two steps employ a plethora of algorithms encompassing various methods to build a 3D point cloud. The aim in the following is to look at these algorithms in terms of applicability and efficiency. \par

\subsubsection{Atom detection\label{atom}}
Atom detection is the extraction of $x$ and $y$ coordinates of each atom from the digital FIM images. A closer inspection of a FIM image reveals that the images are made by a superposition of Gaussian-like distributions of intensities centered around the atomic positions. These atomic positions are also distorted by the image projection from the specimen onto the screen and by the atomic arrangement surrounding the imaged atom. Certain high index planes posses enough corrugation in the electric field so as to enable the ionization of gas atoms almost at every atomic site. The probability of field ionization and hence the intensity of the imaged spot depends on the electric field strength at that atomic site. This explains the intensity differences between the brightly imaged ledge atoms and dimly imaged interior terrace atoms. Dagan's routine \cite{Dagan2017} was based on identifying the atoms which were imaged above a certain threshold intensity. As every atom gets brighter and brighter until it field evaporates, each atom can be detected over the course of the evaporation of a complete plane. This protocol also requires tracking the identified atoms until evaporation. Dagan's routine  is well suited for reconstructing the data to get a statistical distribution of defects and/or ad-atoms. In order to understand the physics of field ionization and field evaporation, as well as to characterize the atomic strains in the imaged lattice, it becomes critical to identify and track every atom right from the moment of its exposure to the surface. In this effort, we developed an alternative method based on edge detection to identify almost all atoms in each image. Two methods for identifying atoms based on edge detection are explained below.

\subsubsection*{Gaussian edge detection}
\begin{figure}
\centering
\includegraphics[width=0.6\linewidth]{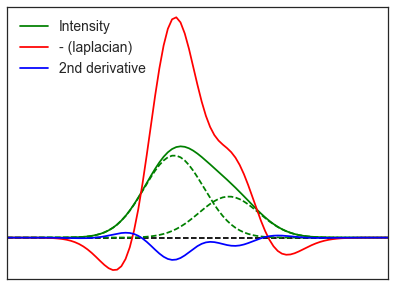}
\caption{A schematic showing the resulting intensity (solid green) of two overlapping Gaussians (dashed green) and the application of a Laplacian operator (red) localizing the two individual contributions. The second derivative (blue) of the Laplacian operator becomes negative in the relevant regions.}
\label{fig:Gaus_laplacian}
\end{figure}

\begin{figure}
\centering
\includegraphics[width=0.9\textwidth]{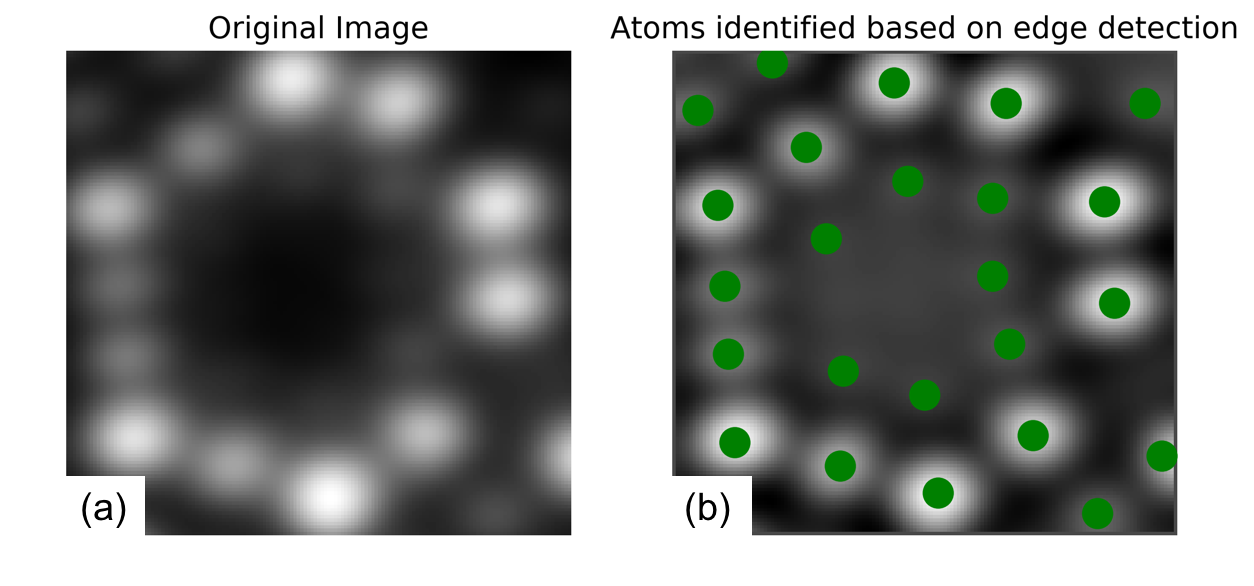}
\caption{Gaussian edge based detection of atoms in FIM image's ROI. Identified atoms shown in (b) from the original image (a) using this procedure are marked in green.}
\label{gaussfim}
\end{figure}

Since the Gaussian-shaped intensity distribution of each atom overlaps with that of the neighboring atoms, detection of individual atoms in a FIM image is not a trivial task. This is particularly the case for atoms from the center of the terrace, where the overlap is much higher as the intensity distribution spreads more widely. To make these intensity distributions more distinguishable from each other, we apply a Laplace operator on the images. As shown in Fig. \ref{fig:Gaus_laplacian} for a one-dimensional Gaussian distribution, the application of a Laplacian operator localizes the intensity distribution for individual Gaussian distributions. In Fig. \ref{fig:Gaus_laplacian} the solid green line is formed by the combination of two individual overlapping Gaussian distributions (the dashed green lines). The red line shows that, by applying the Laplacian operator, the two maxima corresponding to the individual Gaussian intensities (dashed green lines) can now be discriminated as a maximum and a shoulder. These principles are extended to the two dimensional FIM images. Figure \ref{gaussfim}(b) shows the result of applying the Laplacian operator to an original image shown in Fig. \ref{gaussfim}(a). For detecting the atoms, as shown in Fig. \ref{fig:Gaus_laplacian}, the second derivative (blue line) of the localized Laplacian-modified signal can be used as it is more negative in the regions close to the original positions of the two atoms. In a similar way we use the second derivative of the Laplacian-modified image to identify the regions occupied by the atoms and then, the atom's positions are assigned according to the identified region’s centroid. 

\subsubsection*{C spline fitting and peak detection}
Recently, we showed that the local electric field on the atomic sites rearranges over the course of field evaporation giving an impression of atomic movement in FIM images [Katnagallu et al., 2017 communicated]. Another conclusion from this work was that atom positions extracted from less intense parts of the images are least erroneous. Hence for an accurate reconstruction, the information from the center of the terrace needs to be maximized as the corresponding intensity is relatively low. We therefore describe next a method to extract atom positions from the center of a terrace.

The key idea is to improve the edge detection at the center of the terrace by improving the signal to noise ratio (SNR). This is achieved by averaging a certain number of similar images of a given terrace. Provided that the evaporation rate is slow, more images can be recorded for a fixed analysis volume. The images required for this routine are the images of a terrace when it first appears. The image index corresponding to a terrace's first exposure to the surface can be inferred from the analysis shown in Fig. \ref{planeEvap}. The blue circles indicate the images where a new terrace has been exposed to the surface, right after evaporation of the previous terrace. To average a certain number of images with the identified image index, a comparison of similarity for the next few images is performed. The comparison is based on a structure similarity index measure (SSIM) \cite{zhang2011fsim}. The SSIM for two images $i$ and $j$ is given by
\begin{equation}\label{SSIM}
	\textrm{SSIM}(i,j) = [l(i,j)]^{\alpha}\cdot[c(i,j)]^{\beta}\cdot[s(i,j)]^{\gamma},
\end{equation} 
which is a function of the intensity in both images $l(i,j)$ given by the mean of the pixel values, the contrast of both images $c(i,j)$ given by the variance of the pixel values, and the texture of both images $s(i,j)$ given by the covariance of the pixel values in images $i$ and $j$. $\alpha$, $\beta$ and $\gamma$  are weights given to each term. In this case all the weights are set to 1. SSIM is $1$ if the images are identical and is $0$ when the images are completely different. \par

\begin{figure}
\centering
\includegraphics[width=0.9\textwidth]{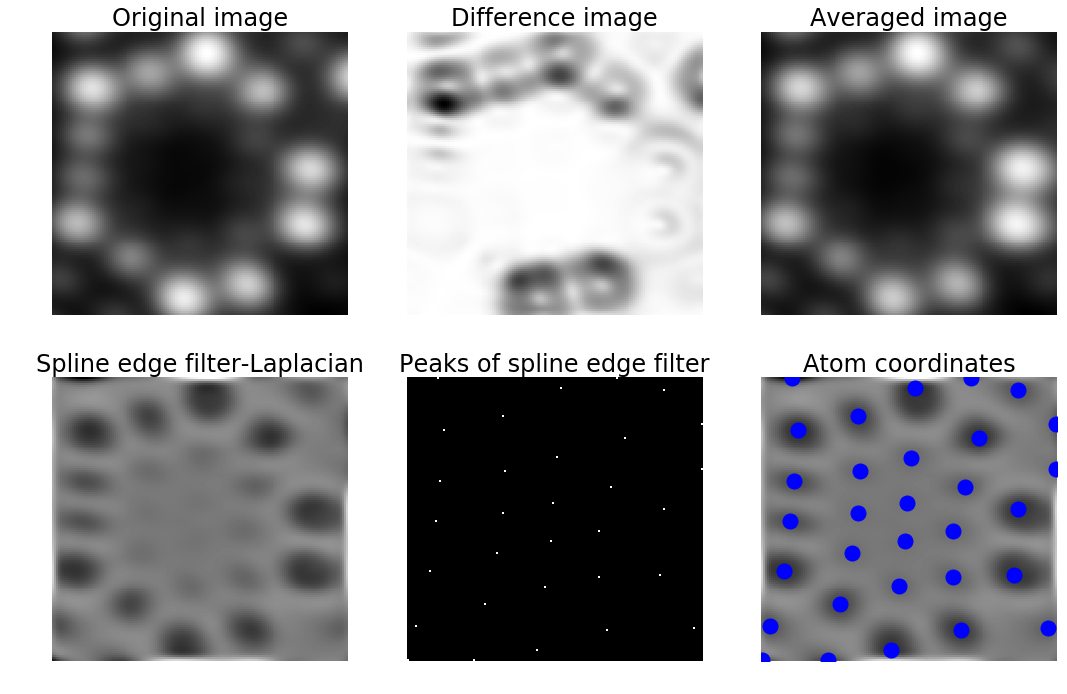}
\caption{C spline fitting and local maxima based detection of atoms in an averaged image. The average image is computed based on the difference image and the spline edge filter and peak based detection are applied to identify the atoms. All atoms are identified due to improved SNR and are marked in blue.}
\label{averaged}
\end{figure}

Thus, the number of images to be averaged is automatically identified based on the computed SSIM. The threshold value was set so that the images to be averaged over should not differ by more than $0.90$ of the SSIM. This value was set based on visual inspection of images where a SSIM of $0.90$ corresponds to one to two atoms being evaporated from the first terrace in the image. Once the number of images to be averaged over is identified, a resultant averaged image is computed. Now the atoms are identified in this image based on a different edge based detection method. First the averaged image is fitted with a cubic spline function in two-dimensions. Using C-spline basis functions to represent a set of samples is advantageous. Operations such as derivation, integration etc., which assume that the data samples are drawn from an underlying continuous function can be computed from the spline coefficients\footnote{https://docs.scipy.org/doc/scipy/reference/tutorial/signal.html}. A Laplacian operator is then convolved with the fitted C-spline function to identify the edges of the given image. The output of the previous operation is used to compute the maxima inside these edges. These local maxima are then recorded as atomic positions. For the case of the tungsten FIM images around the (222) plane where 67 (222) planes were evaporated, the above explained procedure was able to identify 98 $\%$ of atoms. The various steps of the explained method are shown in the Fig. \ref{averaged} for one of the (222) planes.

\subsubsection{Plane determination\label{plane}}
Once the atoms are identified, the detected positions have to be correctly assigned to their corresponding crystallographic planes. As can be seen in Fig. \ref{averaged} (''Original image''), at least two (222) terraces are imaged in the chosen ROI. The outermost terrace is bounded by the brightly imaged atoms and the atoms lying outside this boundary are not of the first terrace. In order to correctly assign the depth coordinate, these atoms need to be classified into those belonging to the top terrace and those not. The methods used to achieve this goal are based on image processing algorithms or on spectral clustering.

\subsubsection*{Image processing based techniques}
\begin{figure}
\centering
\includegraphics[width=0.5\textwidth]{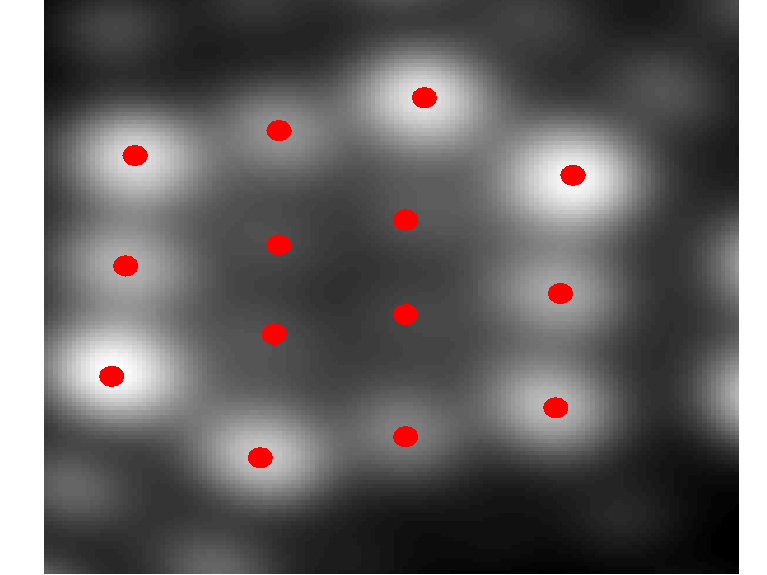}
\caption{A successful application of the entropy based filter to identify the outermost terrace atoms.}
\label{entropy}
\end{figure}

Classifying the atoms according to their crystallographic planes is not trivial, especially using image processing based techniques. Image processing based algorithms rely on identifying the boundaries of the outermost terrace. Since atoms at the ledge of the terrace have a smaller number of nearest neighbors and are the sites of higher electric field, they appear more brightly in FIM. Thus the atoms imaged from the ledge are more similar to themselves than the images of atoms from the terrace's center. This information can be exploited by virtue of the "entropy'' of the image \cite{lewis1990practical}. The entropy, $H$, of the image is given as
\begin{equation}
	H = - \sum_{k} {p_k \log_{2}({p_k})},
\end{equation} 
where $k$ is the number of gray levels and $p_k$ is the probability associated with gray level $k$. Since the atoms at the ledge sites are imaged similarly the pixels in the imaged atoms are associated with a similar $p_k$. An implementation of this entropy as a filter can be used to identify the ledge atoms. Which gives us the boundary for the outermost terrace. The entropy can thus be used as a way to classify atoms to their corresponding terraces. The result of this entropy filter on one of the FIM images is shown in Fig. \ref{entropy}. Other image processing based techniques for this task are watershed segmentation and Fourier filtering based methods. A detailed explanation of these methods is beyond the scope of the current article. 

\subsubsection*{Spectral clustering}

The task of identifying elements in a given dataset based on similarity or dissimilarity amongst them is defined as clustering \cite{Clustering2017}. Clustering is a type of unsupervised learning in which the goal is to partition a set of elements into groups called clusters. Differences between various clustering algorithms are mainly about the ways similarity between the data points is measured \cite{jain1999data}, typically through various distortion or distance measures. We now discuss the so-called spectral clustering applied to plane determination. \par

Eigenvector techniques are frequently employed in multidimensional data in order to extract the underlying correlations of the data (see Sec.~\ref{sec:Dimensionality reduction}). Similarly one can apply such techniques for clustering. Spectral clustering is an Eigenvector based technique. In spectral clustering, we benefit from the node-node adjacency matrix of the graph. For a given graph containing $ N $ nodes (each node corresponds to one point in the dataset), we create a $ N \times N $ adjacency matrix, in which each entry $ (i, j) $ corresponds to the weight of the edge between the nodes $ i $ and $ j $. This essentially corresponds to the similarity between samples $ i $ and $ j $ in the dataset. Such weights $ w_{ij} $ are recorded in a matrix $ W $. As we are working with undirected graphs, the matrix is assumed to be symmetric. It implies that $ w_{ij} = w_{ji} $ $\forall$  $(i, j) $.  Now any clustering algorithm will try to minimize the weights across the clusters. In spectral clustering, the minimization function is constructed based on the adjacency weight matrix and another diagonal matrix called a degree matrix $ D $. Each element in $D$ is the degree of every vertex in the similarity graph, such that  $d_{ii}$ is equal to the sum of the weights of the incident edges, so  $ d_{ii}=\sum\limits_{j=1}^{n}w_{ij} $. \par

In addition, we formally define the Laplacian matrix $ L $ as follows: $ L $ is defined by subtracting the weighted adjacency matrix from the degree matrix. Hence we have $ L= D - W $. The graph Laplacian matrix defined in this way conceals both its structural and its eigenvector behavior. Such a graph Laplacian can now be adopted to identify the pertinent clusters in the data. The number of connected components in the underlying graph can  be related to the number of eigenvectors with zero eigenvalues for the Laplacian matrix $ L $ as they are equal. \par

\begin{figure}
\centering
\includegraphics[width=0.7\linewidth]{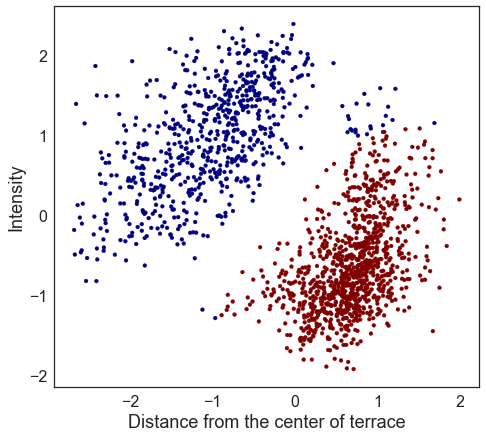}
\caption{Two clusters resulting from spectral clustering indicating two planes present in the given FIM images.}
\label{fig:Spectral_clustering}
\end{figure}

The result of spectral clustering applied to our 3DFIM dataset is shown in Fig. \ref{fig:Spectral_clustering}. As a notion of similarity a combination of two metrics is used. The first metric is the distance between each of the identified atoms, the second is the intensity of the identified atom. Spectral clustering was able to identify two distinct planes in each FIM image as can be seen in Fig. \ref{fig:Spectral_clustering}. Some points near the top right corner were associated to the wrong cluster. This solution to plane determination is still not completely efficient and, here again, the ML could provide significant improvement. The ML algorithms need labeled training data, which will be first labeled by employing the image processing and clustering algorithms outlined herein. The accuracy of these algorithms on each image then needs to be manually verified and that data can be used as a training dataset. \par

\subsubsection{Atom tracking}
Since atoms are imaged multiple times before their evaporation, each atom can be tracked through multiple images. This allows us to identify the position of each imaged atom as a function of its surrounding configuration and also of the operating conditions. In a recent work we used this approach to track the atoms' image positions. We concluded that after field evaporation events there is a local rearrangement of the local electric field, leading to a more precise understanding of how field ion images are formed. Depending on the task and the reconstruction routine used, it might be necessary to track atoms. In the following we describe another clustering algorithm to easily track atoms through multiple images if required.

\subsubsection*{Hierarchal clustering}
\begin{figure}
\centering
\includegraphics[width=0.95\linewidth]{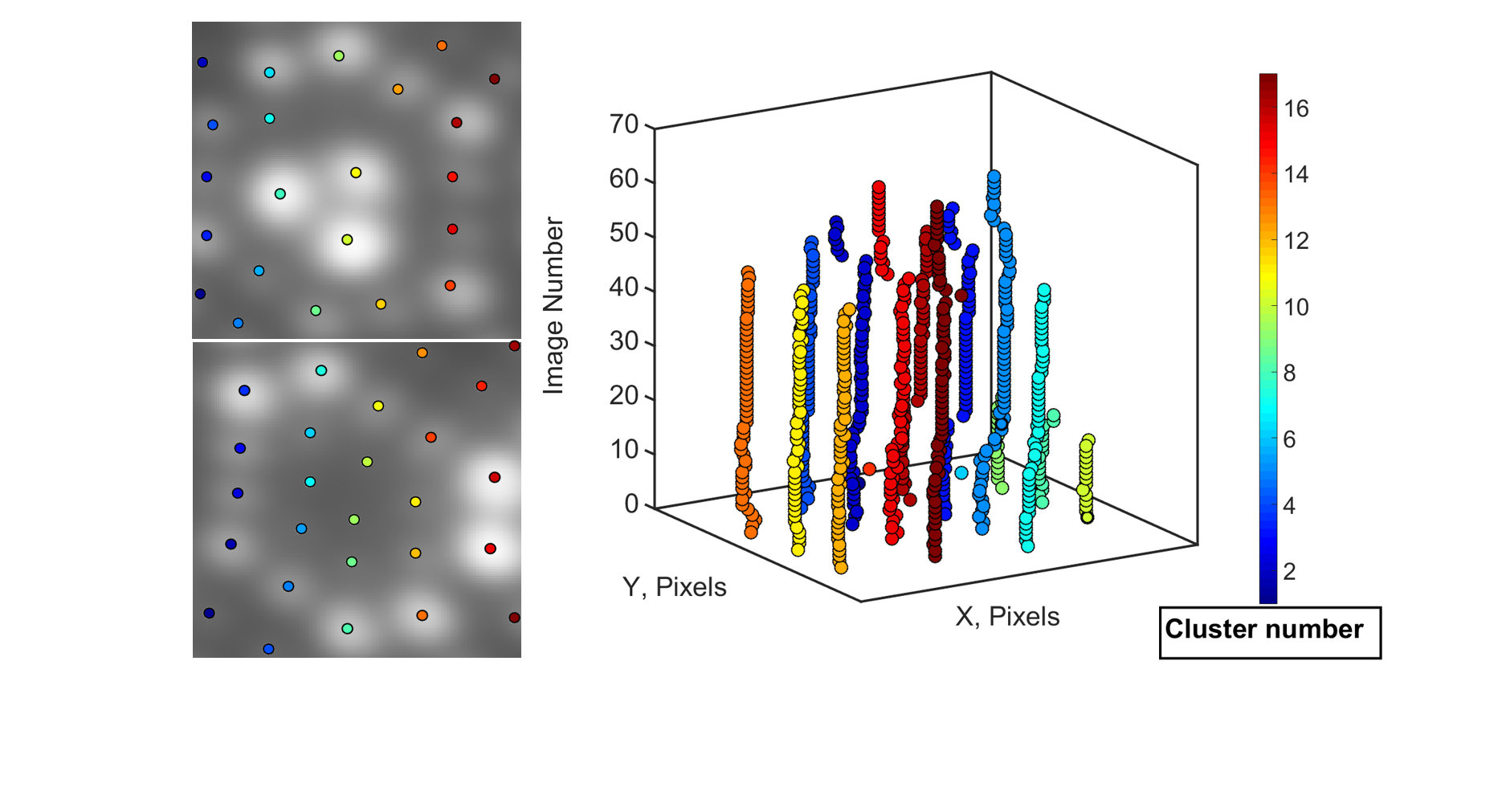}
\caption{Atoms tracked through a sequence of 69 images using hierarchal clustering. Each color indicates the cluster of imaged positions of one atom through the sequence. Images on the left show the start (left bottom) and the end (left top) images of the 69 images sequence.}
\label{Hagg}
\end{figure}

Tracking the same atom imaged multiple times can be considered as a clustering problem. The identified $(x,y)$ coordinates need to be clustered along $z$ such that the final clustering construction is comprised of a set of subsets $K = K_1, K_2,..., K_n$ in $M$, where $M= \bigcup\limits_{i=1}^{n} K_{i}$ and $ K_a \cap K_b = \emptyset $, $\forall$ $a \neq b$  \cite{Rokach2005}. The method used here to cluster these data points is hierarchical clustering \cite{fraley1998many}. In an agglomerative hierarchical clustering, initially every data point is considered as a cluster. These clusters are then merged successively until the sought after cluster construction is found. The merging of clusters is based on a criterion which is a measure of similarity of these clusters. Here, we use a single linkage criterion to connect the clusters. In this criterion, the distance between any two clusters is defined as the shortest distance between any data point of one cluster to any member of the other. The clustering decision is made based on which of the cluster distances is the minimum and those two clusters are merged. The result of using a single linkage hierarchical algorithm on $(x,y)$ coordinates in images corresponding to the evaporation of one (222) terrace is shown in the Fig. \ref{Hagg}. The result is the clustering of theses $(x,y)$ coordinates to form individual atoms tracked through successive images. Each cluster is assigned a number as indicated by the color bar in Fig. \ref{Hagg}. Visual inspection reveals that coordinates corresponding to the same atom through the successive images do not displace by more than 10 pixels. This value was used as a decision metric for the single linkage criterion. The physical grounds for these displacements are discussed in a recent article [Katnagallu et al., 2017 communicated].

\section{A word about 3D field ion microscopy reconstruction}
\begin{figure}
\centering
\includegraphics[width=0.5\linewidth ]{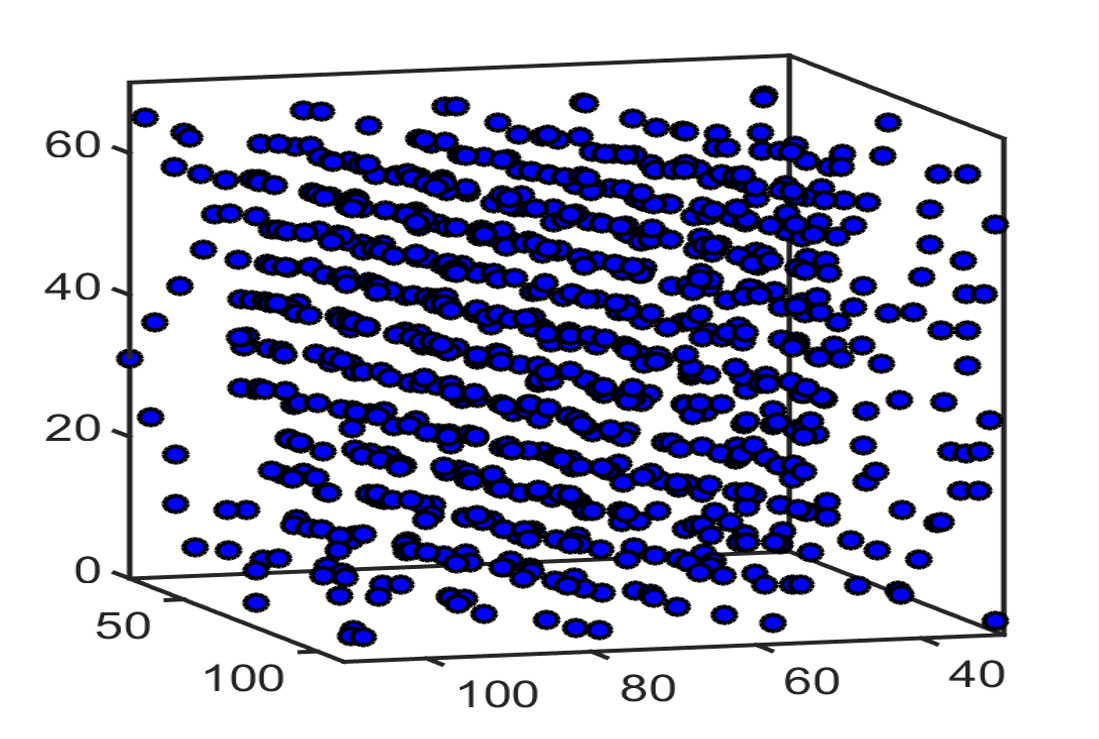}
\caption{Final 3D reconstruction of data extracted from 3DFIM, the dimensions are arbitrary image units.}
\label{3Dview}
\end{figure}

After the data are extracted and labeled correctly, in accordance with previous sections, the data need to be converted to absolute length scales, that are meaningful to materials physics. Previous atom-by-atom approaches, used to calibrate the measured distances between atoms on the image with theoretical nearest neighbor spacing for the crystallographic plane analyzed. Such conversions to an absolute length scale inherently assume that there are no surface relaxations for atoms imaged and atoms are at their equilibrium positions. Figure \ref{3Dview} shows a typical 3D reconstruction of extracted 3DFIM data. The $x$ and $y$ coordinates are in pixels and the $z$ axis is the plane number. The periodicity of the reconstruction is evident as planes in the figure. The noticeable noise in the reconstruction still needs to be understood. One of the sources could be attributed to errors in data extraction. But also, as mentioned before, the image of the local electric field is not directly correlated with the true atomic positions.  Nevertheless the distortion in the reconstructed volume can be expunged by invoking a molecular static relaxation protocol, in which the reconstructed volume is encased in a similarly oriented perfect lattice.  The encasing is done after removing the equivalent atomic sites from the perfect volume. Then using a potential for the material analyzed, a molecular relaxation is done on the encased data in LAMMPS \cite{plimpton2007lammps}. Employing such a routine also preserves the defects recorded in the analyzed volume. The application of this routine on an analyzed 3DFIM volume of W with vacancies can be found in reference \cite{Katnagallu2017}. 

These conversions and relaxations do not affect the statistical distributions of point defects \cite{Dagan2015}, and hence the precision of reconstruction is not a key requirement. On the other hand, if 3DFIM is employed for characterization and quantification of lattice strains, or strains surrounding crystalline defects the precision of the reconstruction needs to be greatly improved. It is important to make a distinction between the resolution of a FIM image and its intrinsic precision. The resolution within a FIM image can be seen as the size of the atom in the image which is determined by the size of the ionization zone, the lateral velocity of the gas atom at the time of ionization and the inherent Heisenberg's uncertainty \cite{Gomer1952,Chen1971}. Whereas the precision of FIM is how well the imaged atom positions correspond to the true atomic sites. Improving the resolution of FIM images can only originate from a reduction in the base temperature of the specimen. However, extracting precise information on the atomic locations requires a thorough comprehension of the image formation in FIM, which directly relates to the details of the complexity of the electric field distribution and its rearrangement when an atom has field evaporated. The presence and the type of defects in the material affects the local bonding and the local arrangement of the atoms on the surface, and hence their effect on the local electric field needs to be precisely quantified. This may be achieved, for instance, by applying ML to the vast amounts of labeled data accessible by 3DFIM. 

 \section{Outlook}
 Field ion microscopy has provided a unique vantage point for atomic scale characterization of materials. The inherent resolution of FIM and the ability to also gather three dimensional information from materials through 3DFIM, puts the technique far ahead of any other contemporary microscopy technique. But the lack of advanced data processing and extraction routines are one of the reasons that likely hindered FIM from becoming a mainstream characterization technique. This article has detailed a burgeoning framework that will allow full data extraction, which we hope will support the renewed interest in FIM and its extension to 3DFIM. The pure image processing techniques have helped us to semi-automate the data extraction from FIM images. The application of machine learning algorithms to our data not only showed the behavior of field evaporation but also helped us to improve the accuracy of the reconstructed data. The layer by layer field evaporation behavior was evident when using PCA or Isomap algorithms on the 3DFIM data. Also using Isomap clustered the images better with respect to number of atoms an image had on its first terrace. The semi automated data extraction routine based on image processing has been helpful not only to improve the data extraction but also helps us to get much more accurately labeled data which can be used as a training dataset for supervised machine learning. With the use of such advanced algorithms for data extraction, we hope not only to completely automate the data extraction from 3DFIM but also identify and characterize various material defects. Apart from data extraction, machine learning has been gaining popularity in identifying fundamental physics phenomena \cite{zdeborova2017machine}. The physics of image formation in FIM is still not completely understood, and it is our firm belief that machine learning can be a powerful tool in this direction.

The discussed methods have been implemented in a set of Python routines. These routines are available upon request to the authors.

\section{Acknowledgments}
SK and BG are grateful to Dr. Michal Dagan, Department of materials, University of Oxford, for providing us with 3DFIM data on tungsten. SK acknowledges IMPRS SurMat for financial support. The funding by the European Research Council (ERC) under the EU’s Horizon 2020 Research and Innovation Programme (Grant No. 639211) is gratefully acknowledged.

\section{References}
\bibliographystyle{ieeetr}
\bibliography{cites.bib,Ex2Sim.bib}

\begin{thebibliography}{10}

\bibitem{muller1955ergebn}
E.~W. M{\"u}ller, ``Ergebn. d. exakten naturwissenschaften. 27 (1953) 330,''
  {\em Z. Elektrochem}, vol.~59, p.~372, 1955.

\bibitem{muller1951Feldionenmikroskop}
E.~W. M{\"u}ller, ``Das {F}eldionenmikroskop,'' {\em Zeitschrift f{\"u}r
  Physik}, vol.~131, no.~1, pp.~136--142, 1951.

\bibitem{Muller1956}
E.~W. M{\"{u}}ller, ``{Resolution of the atomic structure of a metal surface by
  the field ion microscope},'' {\em Journal of Applied Physics}, vol.~27,
  no.~5, pp.~474--476, 1956.

\bibitem{miller2014atom}
M.~K. Miller and R.~G. Forbes, {\em Atom-probe tomography: the local electrode
  atom probe}.
\newblock Springer, 2014.

\bibitem{prosa2017modern}
T.~J. Prosa and D.~J. Larson, ``Modern focused-ion-beam-based site-specific
  specimen preparation for atom probe tomography,'' {\em Microscopy and
  Microanalysis}, vol.~23, no.~2, pp.~194--209, 2017.

\bibitem{de1986best}
C.~de~Castilho and D.~Kingham, ``Best image conditions in field ion
  microscopy,'' {\em Le Journal de Physique Colloques}, vol.~47, no.~C2,
  pp.~C2--23, 1986.

\bibitem{Chen1971a}
Y.~C. Chen and D.~N. Seidman, ``{The field ionization characteristics of
  individual atomic planes},'' {\em Surface Science}, vol.~27, no.~2,
  pp.~231--255, 1971.

\bibitem{Schmidt1993}
W.~A. Schmidt, N.~Ernst, and Y.~Suchorski, ``{Local electric fields at
  individual atomic surface sites: field ion appearance energy measurements},''
  {\em Applied Surface Science}, vol.~67, no.~1-4, pp.~101--110, 1993.

\bibitem{tsong2005atom}
T.~T. Tsong, {\em Atom-probe field ion microscopy: field ion emission, and
  surfaces and interfaces at atomic resolution}.
\newblock Cambridge University Press, 2005.

\bibitem{larson2013local}
D.~J. Larson, T.~J. Prosa, R.~M. Ulfig, B.~P. Geiser, and T.~F. Kelly, ``Local
  electrode atom probe tomography,'' {\em NY: Springer, New York}, 2013.

\bibitem{Brandon1964}
D.~G. Brandon, B.~Ralph, S.~Ranganathan, and M.~S. Wald, ``{A field ion
  microscope study of atomic configuration at grain boundaries},'' {\em Acta
  Metallurgica}, vol.~12, no.~7, pp.~813--821, 1964.

\bibitem{Ranganathan1966}
S.~Ranganathan and J.~B. Thompson, ``{Field Ion Microscopic Observations of
  Dislocation Structures at Grain Boundaries},'' {\em Journal of Applied
  Physics J. Sci. lnstr. J. Appl. Phys. Journal of Applied Physics}, vol.~37,
  no.~12, 1966.

\bibitem{Deconihout1994}
B.~Deconihout, A.~Bostel, P.~Bas, S.~Chambreland, L.~Letellier, F.~Danoix, and
  D.~Blavette, ``{Investigation of some selected metallurgical problems with
  the tomographic atom probe},'' {\em Applied Surface Science}, vol.~76-77,
  pp.~145--154, 1994.

\bibitem{Smith2013}
G.~D.~W. Smith, D.~Hudson, P.~D. Styman, and C.~A. Williams, ``{Studies of
  dislocations by field ion microscopy and atom probe tomography},'' {\em
  Philosophical Magazine}, vol.~93, no.~28-30, pp.~3726--3740, 2013.

\bibitem{Smith1968}
D.~A. Smith, M.~A. Fortes, A.~Kelly, and B.~Ralph, ``{Contrast from stacking
  faults and partial dislocations in field-ion microscope},'' {\em
  Philosophical Magazine}, vol.~17, no.~149, pp.~1065--1077, 1968.

\bibitem{Taunt}
R.~Taunt and B.~Ralph, ``Observations of the fine structure of
  superdislocations in {Ni3Al} by field-ion microscopy,'' {\em Philosophical
  Magazine}, vol.~30, no.~6, pp.~1379--1394, 1974.

\bibitem{Vurpillot2007}
F.~Vurpillot, M.~Gilbert, and B.~Deconihout, ``{Towards the three-dimensional
  field ion microscope},'' {\em Surface and Interface Analysis}, vol.~39,
  pp.~273--277, feb 2007.

\bibitem{Dagan2017}
M.~Dagan, B.~Gault, G.~D.~W. Smith, P.~A.~J. Bagot, and M.~P. Moody,
  ``{Automated Atom-By-Atom Three-Dimensional (3D) Reconstruction of Field Ion
  Microscopy Data},'' {\em Microsc. Microanal}, vol.~23, pp.~255--268, 2017.

\bibitem{Katnagallu2017}
S.~Katnagallu, A.~Nematollahi, M.~Dagan, M.~Moody, B.~Grabowski, B.~Gault,
  D.~Raabe, and J.~Neugebauer, ``High fidelity reconstruction of experimental
  field ion microscopy data by atomic relaxation simulations,'' {\em Microscopy
  and Microanalysis}, vol.~23, no.~S1, pp.~642--643, 2017.

\bibitem{jones2014scipy}
E.~Jones, T.~Oliphant, and P.~Peterson, ``$\{$SciPy$\}$: open source scientific
  tools for $\{$Python$\}$,'' 2014.

\bibitem{vurpillot2017true}
F.~Vurpillot, F.~Danoix, M.~Gilbert, S.~Koelling, M.~Dagan, and D.~N. Seidman,
  ``True atomic-scale imaging in three dimensions: A review of the rebirth of
  field-ion microscopy,'' {\em Microscopy and Microanalysis}, vol.~23, no.~2,
  pp.~210--220, 2017.

\bibitem{Scanlan1971}
R.~M. Scanlan, D.~L. Styris, and D.~N. Seidman, ``{An in-situ field ion
  microscope study of irradiated tungsten: Part II Analysis and
  Interpretation},'' {\em Philosophical Magazine}, vol.~23, no.~186,
  pp.~1459--1478, 1971.

\bibitem{Beavan}
L.~A. Beavan, R.~Scanlan, and D.~Seidman, ``The defect structure of depleted
  zones in irradiated tungsten,'' {\em Acta Metallurgica}, vol.~19, no.~12,
  pp.~1339--1350, 1971.

\bibitem{Dagan2015}
M.~Dagan, L.~R. Hanna, A.~Xu, S.~G. Roberts, G.~D.~W. Smith, B.~Gault, P.~D.
  Edmondson, P.~A.~J. Bagot, and M.~P. Moody, ``{Imaging of radiation damage
  using complementary field ion microscopy and atom probe tomography},'' {\em
  Ultramicroscopy}, vol.~159, pp.~387--394, 2015.

\bibitem{Cerezo1992}
A.~Cerezo, M.~G. Hetherington, J.~M. Hyde, M.~K. Miller, G.~D.~W. {Smith ''},
  and J.~S. Underkoffier, ``{Visualisation of three-dimensional
  microstructures},'' {\em Surface Science}, vol.~266, pp.~471--480, 1992.

\bibitem{Cazottes}
S.~Cazottes, F.~Vurpillot, A.~Fnidiki, D.~Lemarchand, M.~Baricco, and
  F.~Danoix, ``Nanometer scale tomographic investigation of fine scale
  precipitates in a cufeni granular system by three-dimensional field ion
  microscopy,'' {\em Microscopy and Microanalysis}, vol.~18, no.~5,
  pp.~1129--1134, 2012.

\bibitem{Jessner}
P.~Jessner, R.~Danoix, B.~Hannoyer, and F.~Danoix, ``Investigations of the
  nitrided subsurface layers of an fe--cr-model alloy,'' {\em Ultramicroscopy},
  vol.~109, no.~5, pp.~530--534, 2009.

\bibitem{danoix2012atomic}
F.~Danoix, T.~Epicier, F.~Vurpillot, and D.~Blavette, ``Atomic-scale imaging
  and analysis of single layer gp zones in a model steel,'' {\em Journal of
  materials science}, vol.~47, no.~3, pp.~1567--1571, 2012.

\bibitem{Akre2009}
J.~Akr{\'{e}}, F.~Danoix, H.~Leitner, and P.~Auger, ``{The morphology of
  secondary-hardening carbides in a martensitic steel at the peak hardness by
  3DFIM},'' {\em Ultramicroscopy}, vol.~109, no.~5, pp.~518--523, 2009.

\bibitem{mjolsness2001machine}
E.~Mjolsness and D.~DeCoste, ``Machine learning for science: state of the art
  and future prospects,'' {\em Science}, vol.~293, no.~5537, pp.~2051--2055,
  2001.

\bibitem{huang2006kernel}
T.-M. Huang, V.~Kecman, and I.~Kopriva, {\em Kernel based algorithms for mining
  huge data sets}, vol.~1.
\newblock Springer, 2006.

\bibitem{garcia2017review}
A.~Garcia-Garcia, S.~Orts-Escolano, S.~Oprea, V.~Villena-Martinez, and
  J.~Garcia-Rodriguez, ``A review on deep learning techniques applied to
  semantic segmentation,'' {\em arXiv preprint arXiv:1704.06857}, 2017.

\bibitem{drechsler1960analyse}
M.~Drechsler and P.~Wolf, ``Zur analyse von feldionenmikroskop-aufnahmen mit
  atomarer aufl{\"o}sung,'' in {\em Physikalisch-Technischer Teil},
  pp.~835--848, Springer, 1960.

\bibitem{Panayi:2006:3DAP:LAR:FIM}
P.~Panayi, ``A reflectron for use in a three-dimensional atom probe,'' 11 2006.

\bibitem{friedman2001elements}
J.~Friedman, T.~Hastie, and R.~Tibshirani, {\em The elements of statistical
  learning}, vol.~1.
\newblock Springer series in statistics New York, 2001.

\bibitem{lee2007nonlinear}
J.~A. Lee and M.~Verleysen, {\em Nonlinear dimensionality reduction}.
\newblock Springer Science \& Business Media, 2007.

\bibitem{weinberger2006unsupervised}
K.~Q. Weinberger and L.~K. Saul, ``Unsupervised learning of image manifolds by
  semidefinite programming,'' {\em International journal of computer vision},
  vol.~70, no.~1, pp.~77--90, 2006.

\bibitem{Vlachos2017}
M.~Vlachos, {\em Dimensionality Reduction}.
\newblock Boston, MA: Springer US, 2017.

\bibitem{tenenbaum2000global}
J.~B. Tenenbaum, V.~De~Silva, and J.~C. Langford, ``A global geometric
  framework for nonlinear dimensionality reduction,'' {\em science}, vol.~290,
  no.~5500, pp.~2319--2323, 2000.

\bibitem{cox2008multidimensional}
M.~A. Cox and T.~F. Cox, ``Multidimensional scaling,'' {\em Handbook of data
  visualization}, pp.~315--347, 2008.

\bibitem{goodfellow2016autoencoders}
I.~Goodfellow, Y.~Bengio, and A.~Courville, ``Autoencoders,'' in {\em Deep
  learning}, pp.~499--523, MIT Press, 2016.

\bibitem{zhang2011fsim}
L.~Zhang, L.~Zhang, X.~Mou, and D.~Zhang, ``Fsim: A feature similarity index
  for image quality assessment,'' {\em IEEE transactions on Image Processing},
  vol.~20, no.~8, pp.~2378--2386, 2011.

\bibitem{lewis1990practical}
R.~Lewis, {\em Practical digital image processing}.
\newblock Prentice Hall PTR, 1990.

\bibitem{Clustering2017}
C.~Sammut and G.~I. Webb, eds., {\em Clustering}, pp.~226--226.
\newblock Boston, MA: Springer US, 2017.

\bibitem{jain1999data}
A.~K. Jain, M.~N. Murty, and P.~J. Flynn, ``Data clustering: a review,'' {\em
  ACM computing surveys (CSUR)}, vol.~31, no.~3, pp.~264--323, 1999.

\bibitem{Rokach2005}
L.~Rokach and O.~Maimon, {\em Clustering Methods}, pp.~321--352.
\newblock Boston, MA: Springer US, 2005.

\bibitem{fraley1998many}
C.~Fraley and A.~E. Raftery, ``How many clusters? which clustering method?
  answers via model-based cluster analysis,'' {\em The computer journal},
  vol.~41, no.~8, pp.~578--588, 1998.

\bibitem{plimpton2007lammps}
S.~Plimpton, P.~Crozier, and A.~Thompson, ``Lammps-large-scale atomic/molecular
  massively parallel simulator,'' {\em Sandia National Laboratories}, vol.~18,
  2007.

\bibitem{Gomer1952}
R.~Gomer, ``{Velocity Distribution of Electrons in Field Emission. Resolution
  in the Projection Microscope},'' {\em The Journal of Chemical Physics},
  vol.~1772, no.~1952, 1952.

\bibitem{Chen1971}
Y.~C. Chen and D.~N. Seidman, ``{On the atomic resolution of a field ion
  microscope},'' {\em Surface Science}, vol.~26, no.~1, pp.~61--84, 1971.

\bibitem{zdeborova2017machine}
L.~Zdeborov{\'a}, ``Machine learning: New tool in the box,'' {\em Nature
  Physics}, vol.~13, no.~5, pp.~420--421, 2017.

\end{thebibliography}

\end{document}